# Quantum like modelling of the non- separability of voters' preferences in the U.S. political system.

PolinaKhrennikova[a]

[a]School of Management,
University of Leicester, University Road
Leicester LE1 7RH

*pk198@leicester.ac.uk*

**Abstract.** Divided Government is nowadays a common feature of the US political system. The voters can cast partisan ballots for two political powers the executive (Presidential elections) and the legislative (the Congress election). Some recent studies have shown that many voters tend to shape their preferences for the political parties by choosing different parties in these two election contests. This type of behavior referred to by Smith et al. (1999) as "ticket splitting" shows irrationality of behavior (such as preference reversal) from the perspective of traditional decision making theories (Von Neumann and Morgenstern (1953), Savage, (1954)). It has been shown by i.e. Zorn and Smith (2011) and also Khrennikova et al. (2014) that these types of 'non-separable' preferences are context dependent and can be well accommodated in a quantum like framework.

In this paper we use data from Smith et al. (1999) to show first of all probabilistic violation of classical (Kolmogorovian) framework. We proceed with the depiction of our observables (the Congress and the Presidential contexts) with the aid of the quantum probability formula that incorporates the 'contextuality' of the decision making process through the interference term. Statistical data induces the interference term of large magnitude exceeding one (hyperbolic interference). We perform with help of our transition probabilities a state reconstruction of the voters state vectors to test for the applicability of the generalized Born rule. This state can be mathematically represented in the generalized Hilbert space based on hyper-complex numbers.

# 1  Introduction

The application of the quantum models to phenomena outside the remit of quantum physics is not longer perceived as something exotic despite the novelty of this interdisciplinary field. The quantum like models that actively pursue the mathematical framework and concepts of quantum physics to other social phenomena serve as an effective explanatory and descriptive instrument. The philosophy of the application at this stage bears an phenomenological character, without the claim that the social and cognitive systems are exhibiting quantum features ( i.e. that that human brain is a huge quantum system where neurons act like quantum particles).

The domain of applications includes primarily decision making problems in economics and cognitive science[1] (Bruza et al., (2009), Busemeyer et al., (2006a, 2006b, 2007), Franco et al. (2009), Haven and Khrennikov (2009), Photos and Busemeyer (2009), Lambert-Mogilansky and Busemeyer (2012), Asano et al., (2012)). Other applications to decision theory and expected utility violations are by La Mura (2006), Danilov and Lambert Mogiliansky (2006) and Shubik (1999). Many contributions were also achieved in the domain of financial instrument modelling and game theory by Haven (2005, 2006), Segal and Segal (1998), and Eisert et al., (1999)) see also Aerts and Durt (1994) and Aerts and Sozzo (2011) for contributions in logic and categorization. These and many other findings are successfully reviewed in the books of Busemeyer and Bruza (2012) and Haven and Khrennikov (2013).

  A new domain of application is the political science field with first discussions of possibility to model voters' preferences in a quantum framework by Zorn and Smith (2011) and a dynamical representation of the evolution of voters' preferences, including the impact of the so called election campaign "bath", by Khrennikova et al., (2014). Zorn and Smith (2011) fostered for the motivation of the interdisciplinary application of quantum framework:

*Among all of the academic specialties customarily identified as social sciences, political science is perhaps the greatest "debtor" discipline, in the sense that so many of the theories and methods and models put to the task of understanding politics are borrowed from scholars working in other fields.* (Zorn and Smith, (2011), p.83)

In this regard it seemed to be natural to search for inspiration and explanatory power in the domain of quantum physical models and their generalizations. (In this paper we shall consider a generalized quantum model obtained via extension of complex numbers to the hyper-complex algebra.)

---

[1] Many of these findings focus on the violation of classical probabilistic scheme (by Kolmogorov, (1933)) of capturing the events and decision outcomes. The most well known effects that violate the classical representation of observables in the probability space are the conjunction and disjunction errors. These effects emerge due to the presence of contextuality. Its vital impact is measured and discussed in e.g., Conte et al., (2007) Busemeyer and Bruza (2012), Haven and Khrennikov (2009).

## 2   U.S. Political system and the non-separability phenomenon

The US political system has a governance structure based on divided partisan control formed by the so called executive power attributed to the President of the U.S. and the legislative power formed by the Congress of the U.S.[2]. Moreover, two parties historically dominated the political arena: the Democrats and the Republicans. For that reason the U.S. has an established two party political system.

Historically, the voters used to hold stable preferences by supporting the same political party in both the White House and (at least one) Houses of Congress elections. In this regard a large body of orthodox studies on voting preferences perceived such power accumulation as a matter of fact and argued that the divided government should be perceived as a negative occurrence that rather inhibits the normal functioning of the political system (Cutler (1980), McCubbins (1991) Alvarez and Schousen (1993) and others). Zorn and Smith (2011) put forward that voters who strive to maximize their returns in terms of the political power would naturally choose the same party in both types of elections. These types of preferences would be consistent with the postulates of modern decision theories (e.g. Von Neumann and Morgenstern (1953) and Savage (1954)) implying completeness and invariance of the choice alternatives.

However, during the last 40 years the situation with power distribution started to change, in particular after the Watergate scandal related to the presidency of Richard Nixon.[3] The US voters started to search to separate the power as to balance the political decisions.[4] First attention to this phenomenon was paid by Morris Fiorina (1981,1996) who explained this behavior of voters not as a random occurrence, but as a purposive (but not necessarily conscious) motivation to balance the political power as to sustain a less extreme political course (in either direction).

A recent study by Smith et al. (1999) showed that the voters are highly influenced by the upcoming information concerning the outcomes of the Presidential elections. In particular the voters strongly relate the outcomes of the Presidential elections to their subsequent decisions making. In fact they often change their preferences in favor of an opposition party for the Congress elections based on the obtained information. This phenomenon called by Smith et al. (1999) "non- separability" of preferences implies that a particular informational context affects the "mental state" of the voters in a way that is incompatible with their previous beliefs.

More than a half of the respondents in the study by Smith et al (1999) exhibited a "non-separability" of preferences, where the new informational context strongly changed the point of view of the participants. The impact of the new information is especially appearing in the considerable amount of participants that elicited their preferences from uncertainty/ignorance (the answer' don't know) to firm preferences in favor of a particular party in the Congress elections. With line with the psychological studies in decision theory by Kahneman and Tversky (2003) we can witness that the all- inclusive context implying a stability of preferences is not preserved.

---

[2] Senate and the House of Representatives.
[3] For statistical data see e.g., http//www.loc.gov/rr/program/bib/elections/statistics.html.
[4] A type of government where different parties dominate the Congress and the White House is often labelled a "gridlock".

## 2.1     Non- Separability in Quantum framework.

The first endeavor to incorporate the phenomenon of non- separability was done through a static representation of the choice outcomes in a one dimensional Euclidian space (see Fiorina (1996) and Smith et al. (1999)). However, as pointed out by Zorn and Smith (2011) the simplicity of this model entailed some limitations such as a lack of dynamical representation of the decision maker's state evolution as well as no account has been made for additional (contextual) factors impacting the preference emergence, i.e. the upcoming information and the mental characteristics of the voter.

For this reason a Hilbert space representation of the observables (the party preferences) was proposed by depicting the superposition of the initial mental state of the voter. The final state decoherence into the particular eigenvectors is context depended, where a dynamical simulation that also accounted for the impact of the environmental context (the election campaign "bath") was proposed by Khrennikova et al. (2014).

This paper builds on the findings by Khrennikova et al (2014) by showing a violation of classical probabilistic updating of voters beliefs that yields in the violation of the Law of total Probability (Kolmogorov, 1933). We show that the voters' decision probabilities for the decision outcomes exhibit non- additivity, which can be captured through the quantum probability formula (Von Neumann, (1933)). The interference term in this formula accounts for the non- classicality of the observables (we cannot represent them in the classical probability space). We prove that we can apply Born rule by reconstructing the superposition state for our observables from the obtained eigenvalues and the transition probabilities. Furthermore, we obtain such a large magnitude of (called by Khrennikov (2010) "hyperbolic") interference for one of our eigenvectors that constrains a complex number Hilbert space representation of the observables, instead we propose for a hyperbolic Hilbert space state representation.

# 3    Non separability and violation of classical probabilistic framework: empirical evidence.

For the purpose of measuring how the occurrence of non-separability is reflected in the classical probabilistic scheme, we extracted the frequencies from the study of Smith et al. (2009) that presented interview outcomes for 930 respondents[5] before the 1996 elections (with Bob Dole and Bill Clinton as presidential candidates).

The interview had a within group design and was performed in three stages (that we will call "informational contexts") to test for the occurrence of non- separability.

Firstly, a general question was asked to the respondents that sounded: " *Which party is the best choice for managing the U.S. Congress?*" (p.748) The aim of this question was to establish a baseline of the preferences, which subsequently would be compared to the conditioned responses shaped by the next questions. This question was embedded in a context of similar questions about partisan preferences, which assumingly was related to the elimination of order effect.

Secondly, the same respondents were contacted again and provided with following informational context:

"*If Bob Dole were to be elected President, which would you prefer: a Republican Congress to help him pass his agenda or a Democratic Congress to serve as a check on his agenda?*" (Smith et al (1999), p.747). Subsequently, the question was reverted to an opposite informational context (If Bill Clinton were to be elected president…) and posed to the same respondents.

We denote our baseline context as $C = \lambda \{R, D, N\}$, corresponding choice outcomes: Democratic Congress, Republican Congress and 'don't know'.

The contexts for Dole and Clinton are denoted by $P \{R, D\} = \lambda \{R, D, N\}$. Thus we have a baseline context and two conditional observational contexts (Dole/ Clinton) that all can yield three outcomes: republican, democratic and don't' know. We decided to include the participants who were uncertain[6] as well, since the amount of these participants is substantial in the baseline context (as we see from the table 1 it is around 28%).

---

[5] For the baseline question concerning the general preferences for Democratic/Republican congress the sample consisted of 937 respondents.

[6] The ' don't know' answer could be due to uncertainty, but also to a lack of interest to the election contests in general.

We summarize the finding of the interviews in a table below:[7]

|  | C= D | C= R | C=N |
|---|---|---|---|
| Baseline ( C) | 0.323 | 0.406 | 0.271 |
| Dole (P=R) | 0.544 | 0.398 | 0.058 |
| Clinton (P=D) | 0.318 | 0.629 | 0.054 |

**Table 1.** (source: Smith et al., 1999)

### 3.1   Classicality of the obtained probabilities: a check with the Law of Total Probability.

According to De Finetti (1972) the Law of total Probability and the notion of conditional probabilities were applied as the core inputs the of modern decision theory ( in economics and other fields) throughout the 20th Century. The Law of total Probability that is derived with the aid of Bayesian conditional probabilities (see Kolmogorov, (1933)), as is manifest by its name, denotes the total probability of an outcome given its fulfillment through different distinct events.

The formula of the Law of total Probability obeys the principle of additivity and enables to express the occurrence of an outcome (p(C=$\lambda$)) through conditional probabilities (p(C=$\lambda$|P=D), p(C=$\lambda$|P = R) and the marginal probabilities given by p(P=D) and p(P=R). The (P=D) and (P=R) are disjoint events, (P=D) ∩ (P=R) =∅.

It should be denoted that the probabilities applied for statistical decision making (on the right hand of the formula of Total Probability) could be as well objective probabilities, calculated on the basis of the previous statistical experience[8] (e.g. tossing of a coin) as well as subjective probabilities, based on one's personal experience/ beliefs. Both of them are interpretations of the Bayesian probabilities.

In our study we deal with subjective formation of voters preferences, whereas the marginal probabilities encoded in the informational contexts are objective probabilities. Since we have the same amount of study participants (N= 983) in our Dole/ Clinton contexts we treat them as being equal, p(P=D)= p(P=R)= 0.5.

By inserting our frequencies into the Law of total Probability we calculate the baseline probabilities (C=R, C=D and C=N) and examine if they correspond with the obtained results from the baseline context.

$$p(C=D) = (P=D)\ (C=D|P=D) + (P=R)\ (C=D|P=R) \qquad (1)$$

---

[7] The frequencies for conditional probabilities are taken from table 1; p.748 and the compound probabilities are taken from table 2, p. 749. We should note that the baseline context, had N= 937 and Dole / Clinton contexts, N= 983.

[8] We can also express the objective probabilities as given probabilities, which are known to the decision maker from the external sources.

$$0.323 \neq 0.5 * 0.317 + 0.5 * 0.544 = 0.43 \qquad (2)$$

$$p(C = R) = P(P = D)p(C = R|P = D) + p(P = R)(C = R|P = R)$$
$$= 0.406 \neq 0.5 * 0.629 + 0.5 * 0.398 = 0.514 \qquad (3)$$

$$p(C = N) = P(P = D)p(C = N|P = D) + p(P = R)(C = N|P = R) =$$
$$0.271 \neq 0.5 * 0.054 + 0.5 * 0.058 = 0.056 \qquad (4)$$

We obtained violation of the additivity of the conditional probabilities that compose all the baseline outcomes' probabilities. This violation is due to the preference reversal that occurs in the different informational contexts, which changes the mental state of the participants, especially in the 'don't know' outcome. This probabilistic error is well known through the works by Tversky and Shafir (1992) and Croson (1999), when the disjunction of two conditional events is judged as being lower or higher as the compound outcome. We suggest that our observational contexts exhibit non-classicality from the point of view of traditional probability theory and a different state space representation would be necessary to accommodate our observables and their impact on the decision makers' mental states. We proceed with incorporating our probability outcomes in a generalized Law of total Probability, the quantum probability formula that captures the interference of the complex probability amplitudes of the different decision outcomes that are manifest in the superposition state (represented as a state vector). The representation of the observables with help of quantum probability formula is of particular relevance to the field of political science since the so-called " swing voters" that are still in a superposition of various decisions are of particular attention in each US election campaign.

## 4     Generalized quantum representation of belief state

Quantum probability formula is an extension of the classical law of probability with a special interference term; that can exhibit positive, negative or zero interference. Moreover, the particular observational context can yield the traditional 'cos' interference as well as a more exotic hyperbolic interference for which a hyper-complex Hilbert space would be needed.[9]

By applying the quantum probability formula we calculate the interference angles:

---

[9] Were the state vector is represented with hyperbolic numbers $(x + y j, where\ j^2 = +1)$, instead of complex numbers.

$$p(C = D) = p(P = D)p(C = D|P = D) + p(P = R)(C = D|P = R) +$$
$$2\cos\theta\sqrt{(p(P = D)p(C = D|P = D)p(P = R)p(C = D|P = R)}  \quad (5)$$

$$P(C = D) = 0.5 * 0.317 + 0.5 * 0.544 + 2\cos\theta\sqrt{0.5 * 0.317 * 0.5 * 0.544} \quad (6)$$

$$\cos\theta = -0.257;$$
$$\theta = 1.83. \quad (7)$$

This observable outcome yields destructive interference; the absolute value of interference term is less than 1 denoting a trigonometric interference. Therefore it is possible to represent it as cos of some angle (phase). The same appears in the $(C = R)$ outcome:

$$(C = R): \cos\theta = -0.216; \theta = 1.79. \quad (8)$$

The $(C = N)$ case differs crucially from the $(C = D)$ and $(C = R)$ outcomes. Here interference is constructive and the interference term has very high magnitude (which encodes a really strong contextuality effect). Such interference term cannot be represented in the trigonometric form, but in the hyperbolic way as:

$$(C = N): \cosh\theta = 3.84; \theta = 2.021. \quad (9)$$

We can summarize that two of our decision outcomes show destructive interference of sufficiently low (although nonzero) magnitude, a trigonometric interference that can be well explained by the preference reversal were the upcoming information elucidates the subjects' preferences in favor of the respective party in the Congress. Consequently, those participants who do not have any firm preferences in the baseline context exhibit a strongly constructive interference, a hyperbolic interference (for which a hyperbolic Hilbert space representation is needed, see Khrennikov (2010) for a elaboration on how such type of data is modelled and Nyman (2011) for mathematical details.)

When the informational context is altered to (P=D and P=R) the probability of (C=N) is evaporating and the upcoming information is strongly interfering with the initial weak preference attitude. In the next subsection we perform a state reconstruction for the mental state vector of the voters using the obtained probabilities for our outcomes (C=D, C= R and C=N).

### 4.1 Reconstruction of the mental state by the generalized Born rule.

By Born's rule one can reconstruct individual initial mental state Ψ using the matrix of transition probabilities:

$$\begin{matrix} 0.317 & 0.544 \\ 0.629 & 0.398 \\ 0.0539 & 0.058 \end{matrix} \quad (10)$$

Firstly, represent P (C=D) as probability amplitude of Ψ to check if the Born rule (determination of quantum probabilities from probability amplitudes) can be applied:

$$P(C = D) = |\Psi 1|^2 \quad (11)$$

$$\Psi 1 = \sqrt{(P = D)P(C = D|P = D)} + e^{i\theta}\sqrt{P(P = R)\, P(C = D|P = R)}, \quad (12)$$

By Euler's formula we obtain:

$$e^{i\theta} = cos\theta + isin\theta = -0.257 + 0.967i \quad (13)$$

$$\Psi 1 = \sqrt{(0.5 * 0.317)} + (-0.257 + 0.967\, i)\sqrt{(0.5 * 0.544)} = 0.265 + 0.5i = |0.265 + 0.5i| = \sqrt{0.265^2 + 0.5^2} \quad (14)$$

$$|\Psi 1|^2 = 0.32. \quad (15)$$

Next, we reconstruct the state vector for our second outcome in a similar way:

$$P(C = R) = P|\Psi 2|^2 \quad (16)$$

$$\Psi 2 = \sqrt{(0.5 * 0.317)} + (-0.216 + 0.976\, i)\sqrt{(0.5 * 0.398)} = 0.465 + 0.435i$$

$$= |0.465 + 0.435i| = \sqrt{0.465^2 + 0.435^2} \quad (17)$$

$$|\Psi 2|^2 = 0.405 \quad (18)$$

We reconstruct the state vector Ψ3 for the third outcome (C=N) by taking use of the generalized Born rule with hyperbolic numbers:

$$\Psi 3 = \sqrt{(P = D)P(C = N|P = D)} + e^{j\theta}\sqrt{(P = D)P(C = N|P = R)} = \lambda + e^{j\theta}\beta \quad (19)$$

By the hyperbolic analogue of the Euler formula we obtain:
$$e^{j\theta} = cosh\theta + jsinh\theta \quad (20)$$
The elements of the complex hyperbolic algebra have the form:
$$z = x1 + ix2 + jx3; \quad xj \in \mathbb{R},$$
$$\text{where: } |z|^2 = z\bar{z}.$$

We can represent our state vector:
$$\Psi_3 = A + jB = \lambda + e^{j\theta}\beta = (\lambda + \cosh\theta\, \beta) + j\sinh\beta \qquad (21)$$

$$|\Psi_3|^2 = A^2 - B^2 = (\lambda + \cosh\theta\, \beta)^2 - \beta^2 \sinh\theta^2 \qquad (22)$$

We remark that the $\sinh\theta^2$ can be expressed as a relation:
$$\sinh\theta^2 = \cosh\theta^2 - 1 \qquad (23)$$
Thus: $|\Psi_3|^2 = \lambda^2 + 2\lambda\beta\cosh\theta + \beta^2(\cosh\theta^2 - \sinh\theta^2)$
$$= \lambda^2 + 2\lambda\beta\cosh\theta + \beta^2 \qquad (24)$$
By inserting the frequencies from the table 1 we obtain:

$$\Psi_3 = \sqrt{0.5*0.054} + e^{j\theta}\sqrt{0.5*0.058} = 0.164 + 0.17 e^{j\theta} \qquad (25)$$

We express $e^{j\theta}$ through the Euler formula for hyperbolic numbers:

$$\Psi_3 = (0.164 + 0.17\cosh\theta) + 0.17\, j\sinh\theta = 0.81 + 0.17\, j\sinh\theta \qquad (26)$$

$|\Psi^2| = A^2 - B^2 = (0.164 + 0.17\cosh\theta)^2 - 0.029\sinh\theta^2 =$
$0.027 + 2*0.164*0.17*3.84 + 0.029 = 0.27$ [10] $\qquad (27)$

We were able to reconstruct the mental state vector: $\Psi = (\psi_1, \psi_2, \psi_3)$, belonging to the Hilbert space over the hyperbolic algebra. In this space we can select the canonical basis $e_1 = (1,0,0), e_2 = (0,1,0), e_3 = (0,0,1)$. Thus the (hyper-complex) Born rule takes the form:

$$P(D) = |<\psi, e_1>|^2, \qquad P(R) = |<\psi, e_2>|^2, \qquad P(N) = |<\psi, e_3>|^2. \qquad (28)$$

The question observable about Congress elections can be represented as the diagonal operator C in the basis $e_1, e_2,$ and $e_3$. Since the matrix of transition probabilities (10) is not doubly stochastic the questions, about the Presidential elections cannot be represented with the aid of Hermitian operators (in hyper-complex Hilbert space.) One has to use the hyper-complex analog of operator-valued measures.

---

[10] In a hyperbolic space it could be the case that $|z|^2 < 0$ so that negative probabilities could appear. This is not a problem since the negative probabilities are not present as probabilities of the results of mental state measurements. For a detailed treatment of negative probabilities see, e.g., De Barros (2013).

## 5      Concluding remarks

In this article we have examined how the upcoming information concerning the outcomes of election campaign changes the preferences of the voters through analyzing the data from the study by Smith et al (1999). It has been shown that the voters do not only cast ballots for different parties, but even when doing it are highly impacted by the informational context, were the voters with no firm preferences form their opinions by conditioning them on the obtained information. We have shown non-classicality of voters' behavior due to the incompatibility of the observational contexts that cannot be embedded into single classical probability space model.

Based on the obtained probabilities we were able to reconstruct the initial mental state vector of the decision makers that adheres to the (generalized) Born rule. We propose for the representation of the observables that act upon the voters' cognitive states in hyper-complex three-dimensional Hilbert space. Thus our probabilistic representation is not only non-classical (i.e., non-Kolmogorovian), but can neither be described by canonical quantum formalism. Instead its generalization has to be applied.